\documentclass{emulateapj}
\usepackage{amsmath}
\begin{document}

\title{Dark Matter and Synchrotron Emission from Galactic Center Radio Filaments}

\author{Tim Linden\altaffilmark{1,2}, Dan Hooper~\altaffilmark{2,3}, Farhad Yusef-Zadeh\altaffilmark{4} }
\affil{$^1$ Department of Physics, University of California, Santa Cruz, 1156 High Street, Santa Cruz, CA, 95064}
\affil{$^2$Center for Particle Astrophysics, Fermi National Accelerator Laboratory, Batavia, IL 60510}
\affil{$^3$Department of Astronomy and Astrophysics, The University of Chicago, Chicago, IL 60637}
\affil{$^4$Department of Physics and Astronomy, Northwestern University, 2145 Sheridan Road, Evanston, IL 60208}
\shortauthors{}

\begin{abstract}
The inner degrees of the Galactic center contain a large population of filamentary structures observed at radio frequencies. These so-called non-thermal radio filaments (NRFs) trace magnetic field lines and have attracted significant interest due to their hard (S$_v$~$\propto$~$\nu^{~-0.1~\pm~0.4}$) synchrotron emission spectra. The origin of these filaments remains poorly understood. We show that the electrons and positrons created through the annihilations of a relatively light ($\sim$5-10 GeV) dark matter particle with the cross section predicted for a simple thermal relic can provide a compelling match to the intensity, spectral shape, and flux variation of the NRFs. Furthermore, the characteristics of the dark matter particle necessary to explain the synchrotron emission from the NRFs is consistent with those required to explain the excess $\gamma$-ray emission observed from the Galactic center by the Fermi-LAT, as well as the direct detection signals observed by CoGeNT and DAMA/LIBRA. 
\end{abstract}

\slugcomment{FERMILAB-PUB-11-277-A}

\section{Introduction}
Weakly Interacting Massive Particles (WIMPs) provide an attractive class of candidates for the dark matter of our universe~\citep{2005PhR...405..279B, 2009NJPh...11j5006B}. The WIMP paradigm is motivated in part by the realization that particles with weak-scale interactions and masses will naturally freeze-out in the early universe with a relic abundance similar to the observed density of dark matter~\citep{1979ARNPS..29..313S}, an observation referred to as the ``WIMP Miracle". Barring any complicating factors such as conannihilations, resonances, or $S$-wave suppression, dark matter candidates motivated by the WIMP miracle annihilate with a cross section of approximately $\sigma v\approx 3\times 10^{-26}$ cm$^3$~s$^{-1}$ (where $v$ is the relative velocity of the annihilating WIMPs). These annihilations produce stable particles, including photons at $\gamma$-ray energies, as well as protons, electrons, neutrinos, and their antiparticles. Because the dark matter annihilation rate scales with the square of the dark matter density, regions such as the Galactic center, dwarf spheroidal galaxies, and galaxy clusters represent promising locations for searches for indirect signatures of dark matter. 

The launch of the Fermi Large Area Telescope (Fermi-LAT) in 2008 has greatly expanded our view of the $\gamma$-ray sky~\citep{2009ApJ...697.1071A}. In addition to its significantly enhanced effective area, the unparalleled spatial and energy resolution of the Fermi-LAT has allowed for the separation of point sources in the Galactic center, revealing much more detailed information about the diffuse $\gamma$-ray emission from this region \citep{2009ApJS..183...46A, 2009arXiv0912.3828V}. Recently, \citet{2011PhLB..697..412H} identified an excess of $\gamma$-rays within approximately 175~pc of the Galactic center in the energy range of $\sim$500~MeV to 7~GeV, and showed that this could be explained by 7-10 GeV dark matter particles annihilating into $\tau^+\tau^-$ pairs, possibly among other leptonic final states. This range of dark matter masses also provides a suitable match to the direct detection signals observed by the DAMA/LIBRA~\citep{2010EPJC...67...39B, 2010PhRvD..82l3509H} and CoGeNT collaborations~\citep{Aalseth:2011wp, 2010arXiv1002.4703A, Hooper:2011hd}. While upcoming limits from CMB constraints~\citep{2009PhRvD..80d3526S, 2011arXiv1103.2766H} as well as LEP constraints~\citep{2011PhRvD..84a4028F}, error bars in these measurements, along with uncertainties in the dark matter density both locally and near the galactic center, are more than sufficient to remedy any tension between these models. 

If annihilating dark matter particles are in fact responsible for the flux of $\gamma$-rays observed from the Galactic center, this signal is guaranteed to be accompanied by the production of a hard population of electrons and positrons carrying a significant percentage of the total annihilation energy. In the case of democratic annihilation (equal number of annihilations to each family of charged leptons: $e^+ e^-$, $\mu^+ \mu^-$, and $\tau^+ \tau^-$), the combined electrons and positrons carry away nearly an order of magnitude more energy than the $\gamma$-rays observable by the Fermi-LAT (8.4~GeV per annihilation, as compared to 1.1~GeV into $\gamma$-rays for a 8 GeV WIMP). For this reason, the synchrotron radiation from the electrons and positrons produced in dark matter annihilations provides a particularly promising test of the dark matter interpretation of the $\gamma$-ray flux from the Galactic center. Along these lines, it has been shown that the characteristics of the galactic synchrotron excess known as the WMAP Haze are consistent with scenarios in which annihilating dark matter is the source of the observed Galactic center~$\gamma$-rays~\citep{2011PhRvD..83h3517H}. 

In addition to the high dark matter density expected in the Galactic center~\citep{1996ApJ...462..563N}, galactic studies of dark matter synchrotron are enticing due to the many survey observations undertaken over several decades. Observations at 74~MHz~\citep{2003ANS...324...17B}, 330~MHz~\citep{1989ApJ...342..769P, 1991MNRAS.249..262A, 2000AJ....119..207L, 2004AJ....128.1646N}, 1.4~GHz~\citep{2004ApJS..155..421Y}, 5~GHz~\citep{2004ApJ...607..302L} and higher frequencies~\citep{1986PASJ...38..475S, 2000PASJ...52..355R} allow for the additional modeling of spectral features, in addition to the spatial characteristics, of radio sources. A particularly interesting probe for synchrotron signals of dark matter annihilation concerns the population of long ($\sim$40~pc) and thin ($\sim$1~pc) filamentary structures located between 10 and 200~pc from the Galactic center, which have been identified at radio wavelengths~\citep{1984Natur.310..557Y}. These structures, known as non-thermal radio filaments (NRFs), are characterized primarily by their extremely hard radio spectra, with a spectral index $\alpha$ between -0.5 and +0.3 (where S$_\nu$~$\sim$~$\nu^\alpha$), as well as their preferential alignment perpendicular to the Galactic Plane. In several notable cases such as the Radio Arc, high resolution surveys have found a tangled network of separate NRFs which contribute to the overall emission in the structure~\citep{1984Natur.310..557Y}. The radio emission from these filaments is highly polarized, implying that the radio sources are powered by synchrotron emission in a highly ordered poloidal magnetic field of strength $\sim$100 $\mu$G~\citep{1986AJ.....92..818T}.

While several dynamical mechanisms have been proposed to explain the comparatively strong magnetic fields found in NRFs relative to the large scale Galactic magnetic field \citep{1996ApJ...470L..49R, 2006ApJ...637L.101B}, the hard synchrotron spectrum observed from NRFs presents a more difficult puzzle. The observed synchrotron spectral index, $\alpha$, stemming from a power-law electron injection spectrum with index $p$ can be modeled as $p=2\alpha-1$, where $p$ is the power law index of the electron injection spectrum and $\alpha$ is the power law index of the output synchrotron spectrum. Thus, the p~$\sim$~-2.4 spectrum observed to result from astrophysical shock acceleration ~\citep{1987PhR...154....1B, 1991SSRv...58..259J, 2001RPPh...64..429M} would be expected to yield a spectrum of synchrotron emission softer than $\alpha\sim$~-0.7, which is significantly softer than observations of NRFs such as G0.2-0.0, commonly known as the Radio Arc. \citet{1988A&A...200L...9L} modeled the observed synchrotron spectrum of the Radio Arc,  finding the synchrotron spectrum to be best fit by an essentially monoenergetic electron spectrum at approximately 7~GeV. Further observations have found a $\sim$10~GHz turnover in the synchrotron spectrum of many NRFs, implying a strongly peaked electron energy spectrum at several GeV propagating in a magnetic field on the order of 100~$\mu$G~\citep{2006ApJ...637L.101B}.

Several mechanisms have been proposed to explain the origin of such a highly peaked electron spectrum in NRFs. Most notably, \citet{1992A&A...264..493L}, and later \citet{2004A&A...419..161L}, advocated a pile-up scenario in which magnetic reconnection zones formed in collisions between NRFs and molecular clouds create electrical potentials which can accelerate electrons to a single energy specified by the relative strength of the electromagnetic potential and the synchrotron energy loss rate. This explanation, however, is problematic for three reasons: (1) not all bright NRFs have observed molecular cloud associations~\citep{1999ApJ...521L..41L, 2004ApJ...611..858L}, (2) recent simulations of collisional reconnection regimes imply a maximum electron energy of less than 10~MeV, several orders of magnitude below that needed to explain the observed synchrotron signal~\citep{2005MNRAS.358..113L, 2011arXiv1103.5924Z}, although collisionless reconnection, already proposed in the case of $\gamma$-ray bursts, may be significantly more effective in accelerating leptons to high energies~\citep{2010arXiv1011.1904M, 2011P&SS...59..537L}, and (3) the maximum energy of particles moving through magnetic reconnection is expected to depend sensitively on the geometry of the reconnection region~\citep{2011P&SS...59..537L}, which would be expected to produce very different synchrotron emission spectra in different NRFs. An alternate scenario involves electron acceleration in star formation regions~\citep{2003ApJ...598..325Y}, although the necessary mechanism to create a peaked electron spectrum in star formation regions is not known.

In this paper, we propose that dark matter annihilation may provide a physical basis for the nearly monoenergetic electron spectrum necessary to explain NRF emission. Specifically, we find that the highly peaked electron injection spectrum naturally produced by annihilations of light dark matter particles correctly produces the hard and bright synchrotron emission spectrum observed in multiple NRFs. In \S\ \ref{sec:filamentaryarcs}, we review the astrophysics of filimentary arcs and demonstrate that the ordered magnetic field structure within these objects requires that their synchrotron emission results from electrons and positrons injected from within the filaments themselves, rather than from external sources. In \S\ \ref{sec:darkmatter}, we show that relatively light dark matter particles annihilating to leptons, such as those proposed by \citet{2011PhLB..697..412H} to explain the Galactic center $\gamma$-ray excess, are also predicted to inject electrons with a spectrum and intensity naturally capable of explaining the synchrotron intensity and spectrum observed from NRFs. In \S\ \ref{sec:comparison}, we show that synchrotron emission produced as a product of dark matter annihilation can explain the characteristics of the NRF population, including their spectral conformity and the spherical symmetry of their intensity with respect to the Galactic center. Finally, in \S\ \ref{sec:conclusion}, we present several testable predictions for this dark matter annihilation scenario with the goal of discriminating this possibility from astrophysical source mechanisms. Further study of the Milky Way's radio filaments may play a critical role in untangling various interpretations of the $\gamma$-ray excess observed in the Galactic center. 

\section{The Astrophysics of the Non-Thermal Radio Filaments}
\label{sec:filamentaryarcs}
The strength of the magnetic fields in NRFs have been estimated through a variety of means. Early estimates centered on the brightest NRF, G0.2-0.0 (the Radio Arc), and were based on a comparison between the magnetic field pressure and the estimated ram pressure from nearby molecular cloud interactions, which indicated magnetic field strengths as high as 1~mG \citep{1987AJ.....94.1178Y, 1989ApJ...343..703M}. More recent observations, however, have pointed to somewhat weaker magnetic fields ($\sim$~100~$\mu$G) among the population of NRFs for three reasons: (1) the observations of kinks in several NRFs such as G359.1-0.2 (the Snake) and G358.85+0.47 (the Pelican) imply that the magnetic field pressure may be significantly smaller than the ram pressure~\citep{1995ApJ...448..164G, 1999ApJ...521L..41L}, (2) the compact radial extent of the filaments is difficult to explain if a 1~mG magnetic field is surrounded by a region with a significantly weaker galactic magnetic field~\citep{2004ApJ...607..302L}, and (3) synchrotron models of the radio spectrum imply equipartition magnetic fields between approximately 50 and 200~$\mu$G~\citep{1991MNRAS.249..262A}. Furthermore, in the overall population of NRFs, a turnover of the hard synchrotron spectrum at $\sim$10~GHz is observed, implying a magnetic field strength on the order of 100~$\mu$G~\citep{2006ApJ...637L.101B}. 

Although it has been suggested that these relatively strong magnetic fields may simply trace an extremely strong poloidal magnetic field formed in the Galactic center during the proto-halo phase of galaxy formation \citep{1987PASJ...39..843S, 2000ApJ...528..723C, 2010Natur.463...65C}, more recent studies have instead interpreted NRFs to contain local enhancements of the relatively weak diffuse magnetic field \citep[and references therein.]{2006ApJ...637L.101B, 2009A&A...505.1183F}. \citet{2005ApJ...626L..23L} studied the synchrotron spectral index and flux densities at 74 and 330 MHz, and calculated the strength of the large-scale diffuse magnetic field to be approximately $\sim$10~$\mu$G in the inner Milky Way. Furthermore, a ubiquitous poloidal field is unable to explain recent observations of several NRFs which are not aligned perpendicular to the Galactic Disk \citep{1999ApJ...521L..41L, 2004AJ....128.1646N}. Finally, depolarization due to Faraday rotation also implies the strength of the diffuse magnetic field along the line-of-sight to be $\sim$7~$\mu$G~\citep{1995ApJ...448..164G}. 


The strong magnetic fields in NRFs relative to those in the surrounding galactic medium can greatly affect the propagation of cosmic rays into and out of the filaments. The high degree of polarization of the synchrotron emission from NRFs implies that their magnetic fields are highly ordered~\citep{1987AJ.....94.1178Y}. Assuming that depolarization is dominated by turbulence in the magnetic field, the observed fractional polarization can be approximated as:

\begin{equation}
p \approx p_0(\gamma)\frac{B_0^2}{B_0^2+B_r^2} = p_0(\gamma)\frac{B_0^2}{B_T^2}
\end{equation}
where p$_0$($\gamma$) is the intrinsic polarization (determined by the index $\gamma$ of the electron injection spectrum), $B_0$ is the intensity of the ordered field, $B_r$ is the intensity of a random field with variance $\frac{2}{3}B_r$, and $B_T=\sqrt{B_0^2 + B_r^2}$~ is the total field. This solution stands as an equality in the case of a synchrotron spectrum $\alpha$~=~-1, and varies only slightly in all relevant cases~\citep{1966MNRAS.133...67B}. Observations of $\sim$60\% polarization in multiple NRFs \citep{2004ApJS..155..421Y} thus indicate magnetic fields with ordered fractions ($B_0^2/B_T^2$) greater than $\sim$80\%. In reality, depolarizing effects such as Faraday rotation may dominate synchrotron depolarization~\citep{1999ApJ...521L..41L}, implying a magnetic field which is significantly more ordered than required by this lower limit.

\begin{figure}
		\plotone{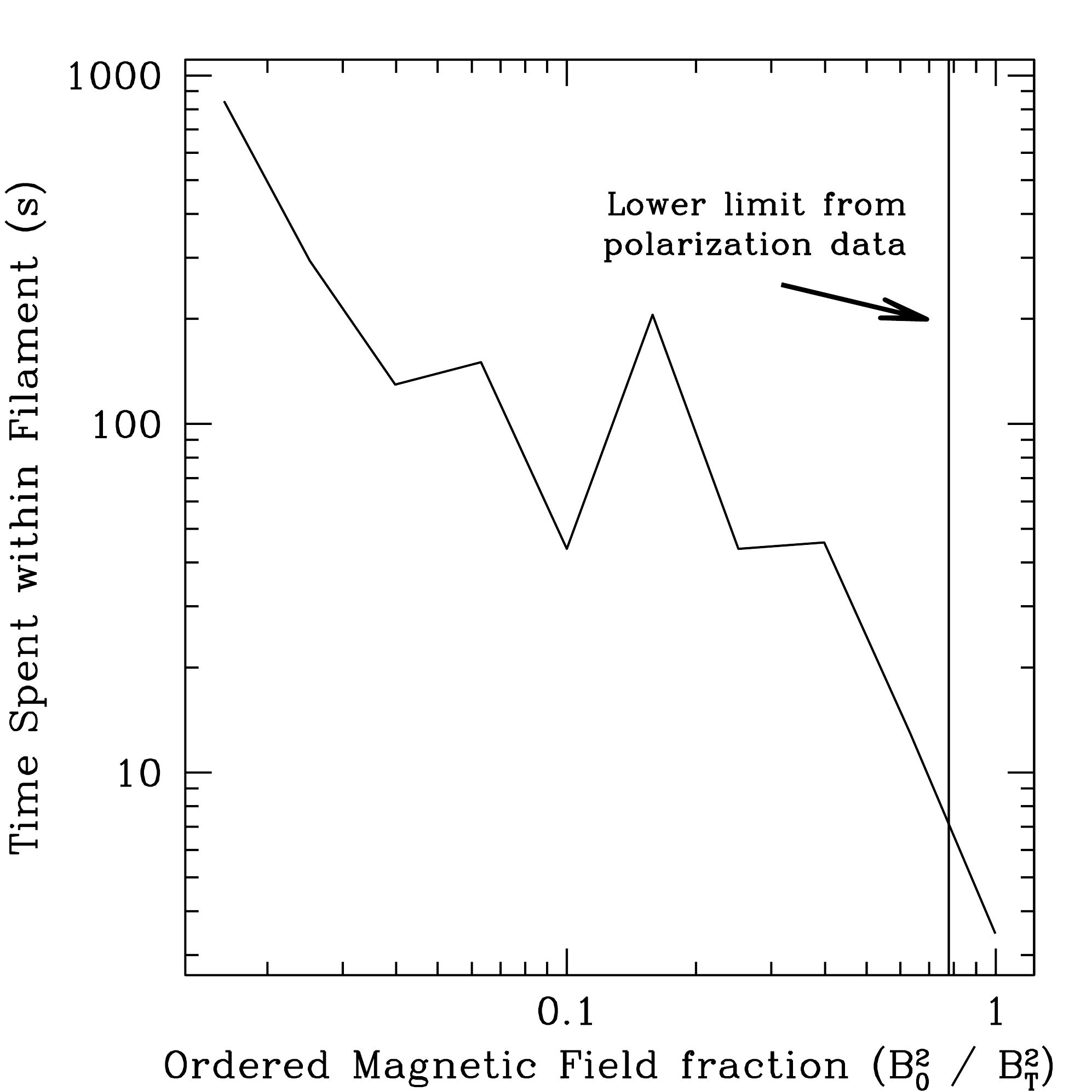}
		\caption{The average time that 10~GeV electrons spend within a filament of total magnetic field of 100 $\mu$G after impacting its side, as a function of the ordered fraction of the magnetic field energy density within the filament. This time is insignificant compared to the $\sim$10$^{12}$ second energy loss time for these particles, indicating that externally produced electrons contribute only a negligible fraction of the synchrotron radiation from NRFs.}
\label{fig:pathlength}
\end{figure}

In the presence of highly ordered magnetic fields, the gyroradius of electrons impinging perpendicular to the field lines of the NRFs is extremely small, leading the filaments to act as a magnetic mirror which quickly rejects incident electrons. In Fig.~\ref{fig:pathlength}, we show the average time that an electron impacting a cylindrical magnetic field with total strength 100~$\mu$G remains within a filament before being evicted, as a function of the ordered fraction of the magnetic field. We employ a filament radius of 0.412 pc (10\"), and imagine the filament as an infinitely long tube in order to safely ignore any edge effects. Using Monte Carlo methods, we have averaged the result over all initial impact angles between the particle and filament. We find that, for the highly ordered magnetic fields required in NRFs, incident electrons are almost immediately expelled from the filament (within seconds) and thus do not significantly contribute to the synchrotron emission from these objects. While electrons may be able to enter through the edges of a filament aligned with the magnetic field, the geometry of NRFs suggests that this would occur for only a very small ($<$~1\%) fraction of electron impacts, greatly diminishing the electron population within the filament. The jaggedness of this dataset is statistical in nature, and is not relevant for the constraints established here.

\section{Synchrotron Emission From Dark Matter Annihilation inside of Non-Thermal Radio Filaments}
\label{sec:darkmatter}

Since magnetic mirroring prevents external electrons from propagating into the NRFs, we are motivated to study the expected synchrotron contribution of electrons and positrons which originate from within the filaments themselves. The injection of energetic electrons/positrons through dark matter annihilations is particularly interesting for four reasons: (1) the dark matter annihilation rate will remain entirely unaffected by the peculiar magnetic properties of the NRFs, allowing for a rigorous comparison in flux between the filaments and the surrounding interstellar medium, (2) the electron injection spectrum from dark matter will be identical at all locations, providing an explanation for the similar synchrotron spectra observed from multiple NRFs, (3) the dark matter density in the Galactic center is dominated by a smooth component \citep{2008Natur.454..735D}, and naturally explains the intensity observed from all NRFs, and (4) a dark matter annihilation scenario at very similar energies has already been proposed to explain Galactic center emission at $\gamma$-ray energies~\citep{2011PhLB..697..412H}.

The local dark matter annihilation rate per volume at a location $\vec x$ is given by:

\begin{equation}
\Phi_{DM} (\vec x)~=~\frac{1}{2}\langle\sigma v\rangle~\Bigg(\frac{\rho(\vec x)}{M_{DM}}\Bigg)^2
\end{equation}

\noindent where $\rho$($\vec x$) is the dark matter energy density, $M_{DM}$ is the dark matter particle's mass, and $\langle\sigma v \rangle$ is dark matter's annihilation cross section multiplied by the relative velocity of the WIMPs. We adopt a dark matter distribution based on the results of numerical simulations~\citep{1996ApJ...462..563N, 2008Natur.454..735D}, and given by $\rho(r) = \rho_0 \, (r/8.5 \, {\rm kpc})^{-1.25}$, where $r$ is the distance from the Galactic center and $\rho_0$~=~ 0.385~GeV~cm$^{-3}$~\citep{2010JCAP...08..004C, 2011arXiv1105.4166L}. We also adopt a dark matter annihilation cross section of $\langle\sigma v\rangle=3\times~10^{-26}$~cm$^3$~s$^{-1}$, which is the value predicted for a simple thermal relic. 

Assuming approximate cylindrical symmetry for the filament geometry, the annihilation rate within a filament of length $l$ and diameter $w$ is given by:

\begin{equation}
\label{eq:annihilation_rate}
\begin{split}
\Phi_{DM}~=1.4\times10^{33}~{\rm s}^{-1}~\Bigg(\frac{8~{\rm GeV}}{M_{DM}}\Bigg)^2~\Bigg(\frac{\langle\sigma v\rangle}{3\times 10^{-26} {\rm cm}^3/{\rm s}}\Bigg)\\  
\times~\Bigg(\frac{r}{100\, {\rm pc}}\Bigg)^{-2.5}~\Bigg(\frac{l}{40\,{\rm pc}}\Bigg)~\Bigg(\frac{w}{1\,{\rm pc}}\Bigg)^2~~~~
\end{split}
\end{equation}

\noindent where $r$ is the distance of the filament from the Galactic center. We note that in many NRFs, the distance from the Galactic center changes considerably across the length of the filaments - we will discuss the effect of this when modeling specific filaments. 

The types and spectra of particles produced in dark matter annihilations depend on the details of the particle physics model. In order to generate a bright flux of synchrotron emission with a spectrum peaking at $\sim$10 GHz, we will focus on dark matter which annihilates dominantly to charged leptons. In particular, we will consider a democratic model which annihilates equally to $e^\pm$, $\mu^\pm$, and $\tau^\pm$ final states. The nearly instantaneous decays of the taus and muons produce lower energy electrons/positrons, as well as a prompt flux of $\gamma$-rays (as opposed to $\gamma$-rays from the inverse-Compton scattering of energetic electrons). It is this prompt flux of $\gamma$-rays which \citet{2011PhLB..697..412H} find to be consistent with the excess observed in the Galactic center by Fermi-LAT.

\begin{figure}
		\plotone{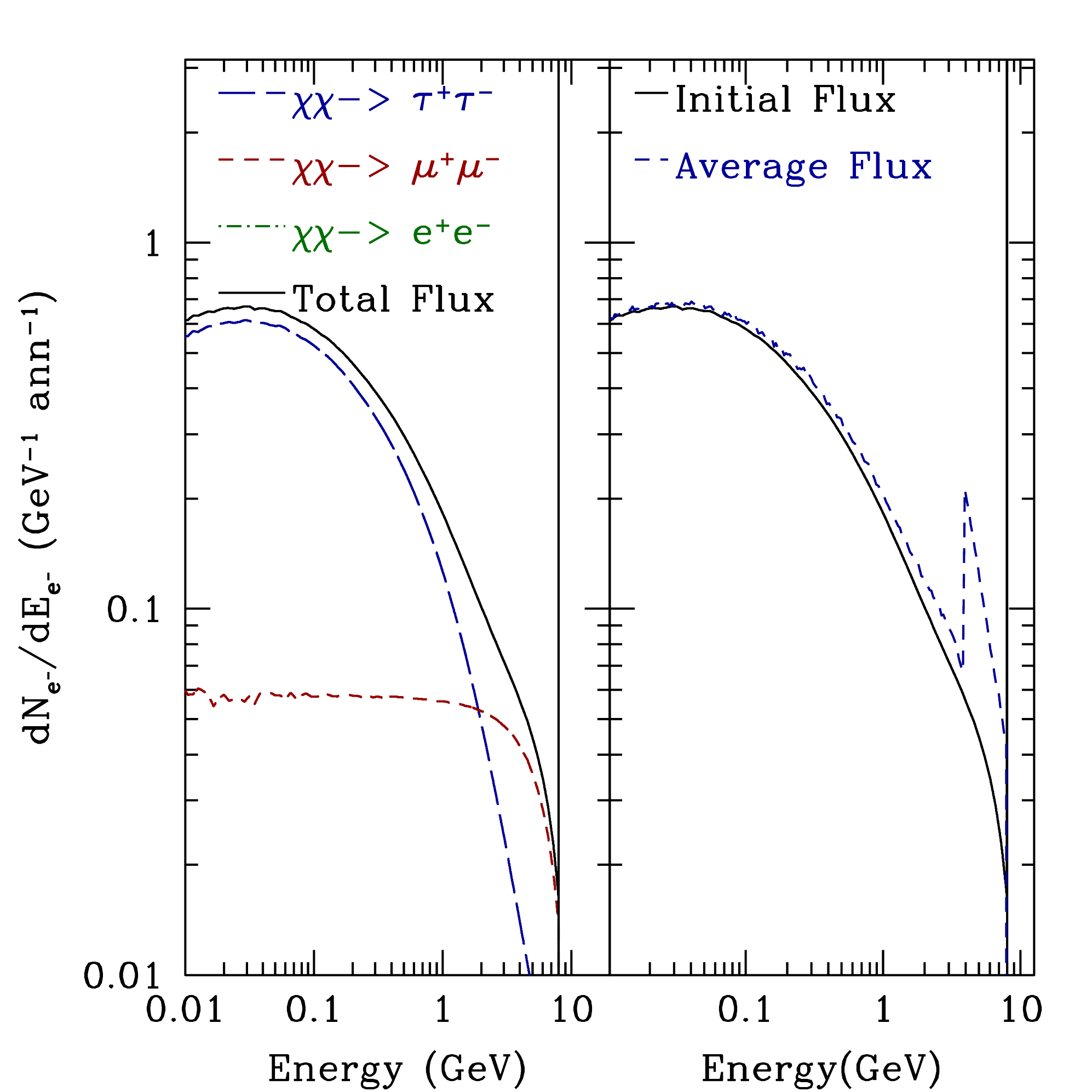}
		\caption{(Left) The spectrum of electrons injected through the annihilation of an 8~GeV dark matter particle to $e^+ e^-$, $\mu^+ \mu^-$ and $\tau^+ \tau^-$ (democratically). Note that the contribution from $e^+ e^-$ takes the form of a delta function at 8 GeV and is concealed by the line denoting the ``Total Flux''. (Right) The spectrum of electrons from dark matter annihilations before (solid) and after (dashed) synchrotron energy losses for an energy loss time of $\tau$~=~1.0 (as defined in Eqn.~\ref{eq:synch_time}). The direct flux to electrons represents a delta function at the 8~GeV mass of the dark matter particle which carries approximately 2/3 of the total electron energy. The positron flux has an identical energy dependence and lends a factor of two to the total lepton flux from dark matter annihilation.}
\label{fig:leptonflux}
\end{figure}

The spectrum of electrons and positrons produced through dark matter annihilations within a given NRF is calculated using the Pythia package~\citep{2001CoPhC.135..238S, 2004JCAP...07..008G}. In the left frame of Fig.~\ref{fig:leptonflux}, we show the injected electron spectrum per dark matter annihilation for our canonical case of a dark matter particle with a mass of 8~GeV, annihilating equally into $e^+ e^-$, $\mu^+ \mu^-$ and $\tau^+ \tau^-$. We note that the electron/positron spectrum is very hard, following a spectrum between E$^{-0.5}$ and E$^{0}$ between 100 MeV and 8 GeV, although we caution that this spectrum is not a continuous power law. The majority ($\sim$2/3) of the electron energy is deposited in a delta function at 8~GeV following the dark matter annihilations directly into electrons. These 8~GeV electrons will dominate the synchrotron spectrum from NRFs, in part due to their shorter synchrotron energy loss time. We note that the positron spectrum is identical and lends another factor of two to the overall synchrotron flux.

In order to determine the synchrotron spectrum expected from dark matter annihilations, we must also model the diffusion of electrons throughout the NRFs. Models of the filaments typically require an electron diffusion timescale similar to the energy loss time of the electron population in the NRF's magnetic field~\citep{1995ApJ...448..164G, 1999ApJ...526..727L}. This has the effect of smearing out the electron energy distribution and softening the overall synchrotron spectrum. Through considerations of the electron gyroradius similar to those discussed in \S\ \ref{sec:filamentaryarcs}, electrons created within the NRFs are constrained from effective diffusion perpendicular to the ordered magnetic field. In the case of an entirely ordered magnetic field, charged leptons would spiral freely along the magnetic field lines until exiting the filament. However, in the observed regime containing significant ordered and unordered fields, diffusion is expected to be significantly more complicated. 

In the case of very low turbulence levels, much work has been done within the perturbative framework of quasi-linear theory which seeks to calculate the parallel and perpendicular diffusion components as a function of the power in turbulent modes of the magnetic field with a wavenumber resonant with the inverse of the particles momentum~\citep{1966ApJ...146..480J, 1971RvGSP...9...27J}. The amplitude of these modes, however, is poorly constrained in galactic simulations. More recently, numerical simulations have been used to analyze the parallel and perpendicular diffusion constants in regimes in which the ordered and unordered field are co-dominant. Notably, \citet{2002PhRvD..65b3002C} found that in the case of a magnetic field which is approximately 80\% ordered, the parallel diffusion constant exceeds the perpendicular diffusion constant by a factor of $\sim$125. Very similar results were later obtained for the case of more energetic cosmic rays~\citep{2007JCAP...06..027D}. Since the length travelled by diffusive particles can be written as $\ell$~=~$\sqrt{2Dt}$, where D is the assumed diffusion constant. This implies that perpendicular and parallel diffusion will remove particles from the filaments on equivalent timescales if the parallel and perpendicular extent of the NRFs is approximately 125 to 1. This length-scale is similar to the observed length to width ratio in multiple filamentary structures \citep{2004ApJS..155..421Y}, implying that diffusion along the length of the filamentary arcs acts on a similar time frame to diffusion across the much smaller NRF width. Since the calculation of magnetic field order stands as a lower limit in this calculation, it is feasible that perpendicular diffusion is in fact entirely irrelevant in the population of NRF structures. 

The overall normalization of the diffusion constant depends sensitively on the length scale of the turbulent disturbances in the magnetic medium~\citep{1966ApJ...146..480J} and is highly uncertain. While simulations are able to constrain the mean galactic diffusion constant through observations of cosmic ray primary-to-secondary ratios at the solar position~\citep[e.g][]{1998ApJ...509..212S}, these simulations do not constrain local diffusion constants, especially in magnetically unique regions such as NRFs. 

The synchrotron energy loss time of an electron is given by:

\begin{equation}
\label{eq:synch_time}
\frac{E}{\dot E}~=~6.6 \times 10^{12}~ {\rm s}\,~\Bigg(\frac{8 \,{\rm GeV}}{E}\Bigg)~\Bigg(\frac{100~\mu G}{B}\Bigg)^2
\end{equation}

Due to the difficulties of calculating the diffusion constant within a partially ordered magnetic field, we choose a parallel diffusion constant such that electrons remain within the NRF for a length of time given by:

\begin{equation}
\label{eq:synch_time}
\begin{split}
T_{\rm confinement} =\frac{E}{\dot E} \, \times \, \Bigg(\frac{8 \, {\rm GeV}}{E}\Bigg)^{0.33} ~~~~~~~~~~~~~~~~~~~~~~~~ \\ ~~=~6.6 \times 10^{12}~ {\rm s}\,~\Bigg(\frac{\tau}{1}\Bigg)~\Bigg(\frac{8 \,{\rm GeV}}{E}\Bigg)^{1.33}\Bigg(\frac{100~\mu G}{B}\Bigg)^2
\end{split}
\end{equation}
where $\tau$ is the ratio of the diffusion timescale for 8~GeV electrons compared to their synchrotron energy loss time. For example, in the case $\tau$~=~1.0, 8~GeV electrons diffuse out of the NRF on a timescale equal to their synchrotron energy loss time. Thus the $\tau$ parameter can also be seen as an indicator of the average synchrotron exhaustion, or the average time that an electron generated by dark matter annihilation has propagated through the NRF before producing the synchrotron emission presently observed. The additional factor of E$^{-0.33}$ accounts for the energy dependence of the diffusion constant calculated by~\citet{1941DoSSR..30..301K}. We note that the synchrotron softening of an electron spectrum for a given value of $\tau$ is independent of the magnetic field strength. In Fig.~\ref{fig:leptonflux} (right), we show the spectrum of electrons from dark matter annihilations after accounting for synchrotron energy losses for $\tau$~=~1.0.

\begin{figure}
		\plotone{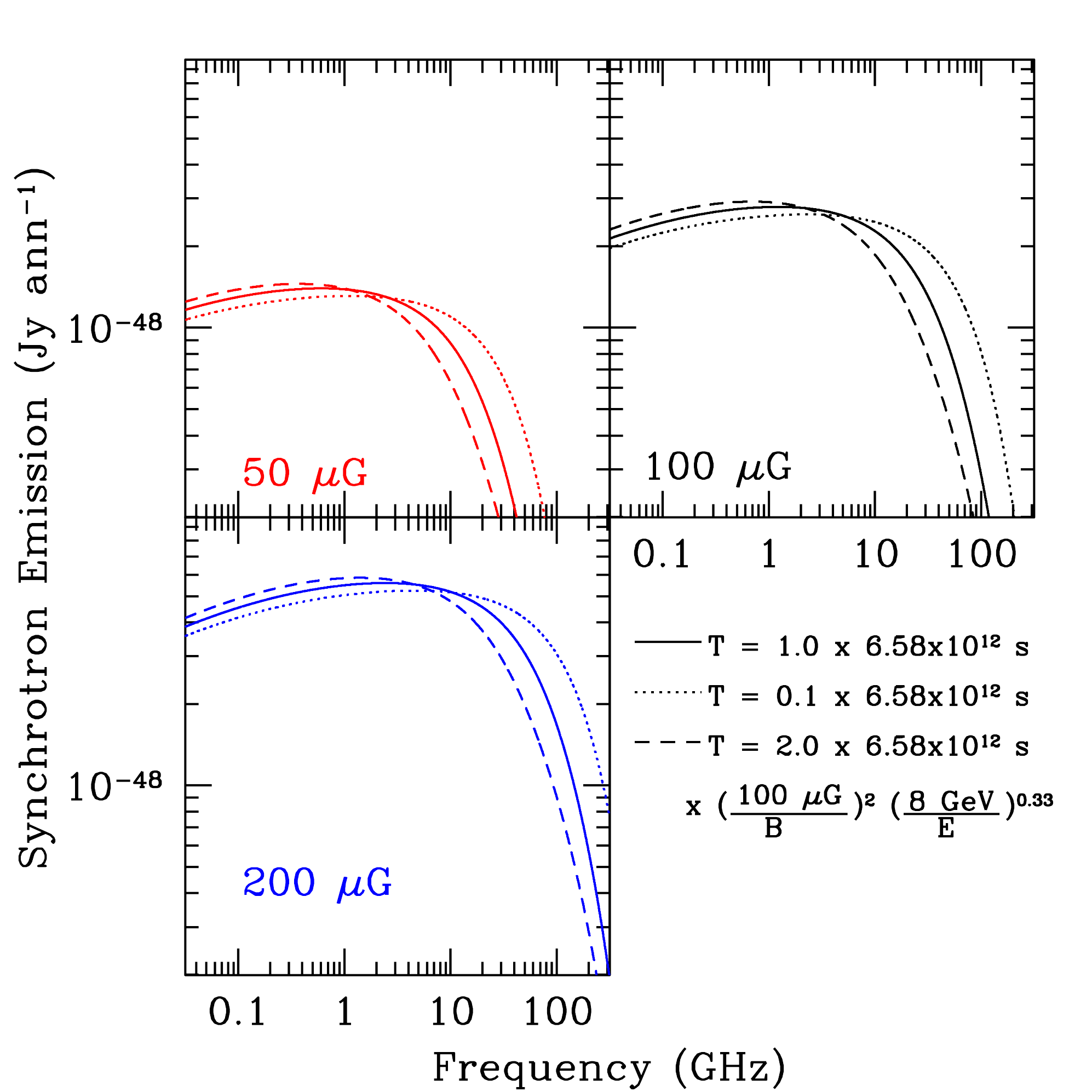}
		\caption{The spectrum of synchrotron radiation (in Janskys, defined as $10^{-26}$ watts per square meter per Hz) from electrons produced by the annihilation of an 8~GeV dark matter particle democratically into leptonic final states in magnetic fields of 50~$\mu$G (red, top left), 100 $\mu$G (black, top right), and 200 $\mu$G (blue, bottom), with the electron distribution softened during propagation times of $\tau$~=~1.0 (solid), 0.1 (dotted) and 2.0 (dashed).}
\label{fig:synchrotron_spectrum}
\end{figure}

We are now prepared to calculate the synchrotron spectrum resulting from dark matter annihilations taking place within a NRF. In Fig.~\ref{fig:synchrotron_spectrum}, we plot the synchrotron spectrum from dark matter annihilations for magnetic field strengths of 50~$\mu$G, 100~$\mu$G, and 200$\mu$G and for values $\tau$~=0.1, 1.0 and 2.0. In each case, we predict a peak in synchrotron energy at $\sim$1-10~GHz followed by a suppression of the synchrotron emissivity at higher frequencies. In the following section, we will compare this prediction to the synchrotron spectrum observed from specific NRFs.

\section{Comparison to Specific Filaments}
\label{sec:comparison}


In astrophysical interpretations of NRF observations, variations in the electron injection spectrum can be invoked to effectively explain the different spectral features in each NRF, since the peak of the synchrotron emission spectrum depends on the square of the electron energy. However, in the case of dark matter annihilations, the injected electron spectrum must be uniform in each filament. Variations in the observed synchrotron spectra may still originate from differences in either the magnetic field strength or diffusion timescales of each NRF. These effects are relatively weak, however, and would be unable to explain extreme variations in the spectral turnover of different NRFs. Thus a population survey of the synchrotron spectra in NRFs remains a powerful diagnostic for testing the dark matter interpretation. 
 
\begin{deluxetable*}{cccccc}
\label{tab:spectra}
\tabletypesize{\scriptsize}
\tablecaption{\label{tab:spectra}Spectral Characteristics of Observed Non-Thermal Radio Filaments}
\tablewidth{0pt}
\tablehead{
\colhead{Name} & \colhead{Alternative Name} & \colhead{$\alpha^{1.4 GHz}_{0.33 GHz}$} & \colhead{$\alpha^{4.8 GHz}_{1.4 GHz}$}& \colhead{$\alpha^{>}_{4.8 GHz}$} & \colhead{References}
}
\startdata
G0.08+0.15 & Northern Thread & -0.5 & -0.5 & -2.0 & \citet{1999ApJ...526..727L, 2000AJ....119..207L}\\
G358.85+0.47 & The Pelican & -0.6 & -0.8 $\pm$ 0.2 & -1.5 $\pm$ 0.3 & \citet{1999ASPC..186..403K, 1999ApJ...521L..41L}\\
G359.1-0.2 & The Snake & -1.1 & $\sim$0.0 & * & \citet{1993PASAu..10..233N, 1995ApJ...448..164G}\\
G0.2-0.0 & Radio Arc & ------- & +0.3 & +0.3 & \citet{1992PASJ...44..367S, reich2003}\\
G0.16-0.14 & Arc Filament &  ------- & 0.0 & -0.8 & \citet{1992PASJ...44..367S}\\
G359.32-0.16 & ------- & -0.1 & -1.0 & ----------- & \citet{2004ApJ...607..302L}\\
G359.79+0.17 & RF-N8 & -0.6 $\pm$ 0.1 & -0.9 to -1.3 & ---------- & \citet{2008ApJS..177..515L} \\
G359.85+0.39 & RF-N10 & 0.15 to -1.1$^{**}$ & -0.6 to -1.5$^{**}$ & ---------- & \citet{2001ApJ...563..163L, 2008ApJS..177..515L} \\
G359.96+0.09 & Southern Thread & -0.5 & ---------- & ----------- & \citet{2000AJ....119..207L}\\
G359.45-0.040 & Sgr C Filament & -0.5 & ------------- & -0.46 $\pm$ 0.32 & \citet{1995ApJS...98..259L, 2008ApJS..177..515L} \\  
G359.54+0.18 & Ripple & --------- & -0.5 to -0.8 & ---------- & \citet{2008ApJS..177..515L} \\
G359.36+0.10 & RF-C12 & --------- & -0.5 to -1.8 & ---------- & \citet{2008ApJS..177..515L} \\
G0.15+0.23 & RF-N1& --------- & +0.2 to -0.5 & ---------- & \citet{2008ApJS..177..515L} \\
G0.09-0.09 & --------- & --------- & ---------- & 0.15 &  \citet{reich2003} \\
\enddata 
\tablenotetext{*}{Two very different values exist in the literature for the high frequency spectrum of G359.1-0.2. \citet{1995ApJ...448..164G} cites a value of -0.2 $\pm$ 0.2, while a more recent analysis by \citet{2008ApJS..177..255L} yields $\alpha^{8.33}_{4.8 GHz}$~=~-1.86 $\pm$ 0.64}
\tablenotetext{**}{Spectrum is highly position dependent, but shows a clear trend towards steeper spectral slopes at high frequencies for any given position}
\end{deluxetable*}

In Table~\ref{tab:spectra}, we have compiled the observed synchrotron spectra of the most thoroughly studied NRFs. We find the population to be relatively homogeneous, with a hard spectrum below $\sim$5~GHz that quickly turns over at higher frequencies. The variation in the spectral turnover from the hardest NRF (G0.2-0.0, Radio Arc) to the softest (G0.08+0.15, Northern Thread) is approximately an order of magnitude.

In order to test whether magnetic field and diffusion timescale variations can explain these spectral and intensity variations within the highly constrained framework of a uniform electron injection spectrum, we consider four NRFs with particularly well measured spectra and intensities: G0.2-0.0 \citep[the Radio Arc,][]{reich2003}, G0.16-0.14~\citep[the Arc Filament,][]{1992PASJ...44..367S}, G0.08+0.15~\citep[the Northern Thread,][]{1999ApJ...526..727L} and G359.1-0.2~\citep[the Snake,][]{1995ApJ...448..164G}. Data were extracted using the Dexter package~\citep{2001ASPC..238..321D}, and is shown with the statistical errors and astrophysical background subtraction determined by each study. In each case, we calculate the flux at the point of maximum 1.4~GHz emission, with the exception of the Snake. For that filament, the peak emission corresponds to a ``kink" in the NRF morphology which shows a spectral index representative of astrophysical injection. Thus for the Snake, the flux is determined at a point where the spectral index $\alpha^{4.8~GHz}_{1.4~GHz}$ is entirely flat, which lies at approximately 19' in the \citet{1995ApJ...448..164G} nomenclature.

\begin{figure}
		\plotone{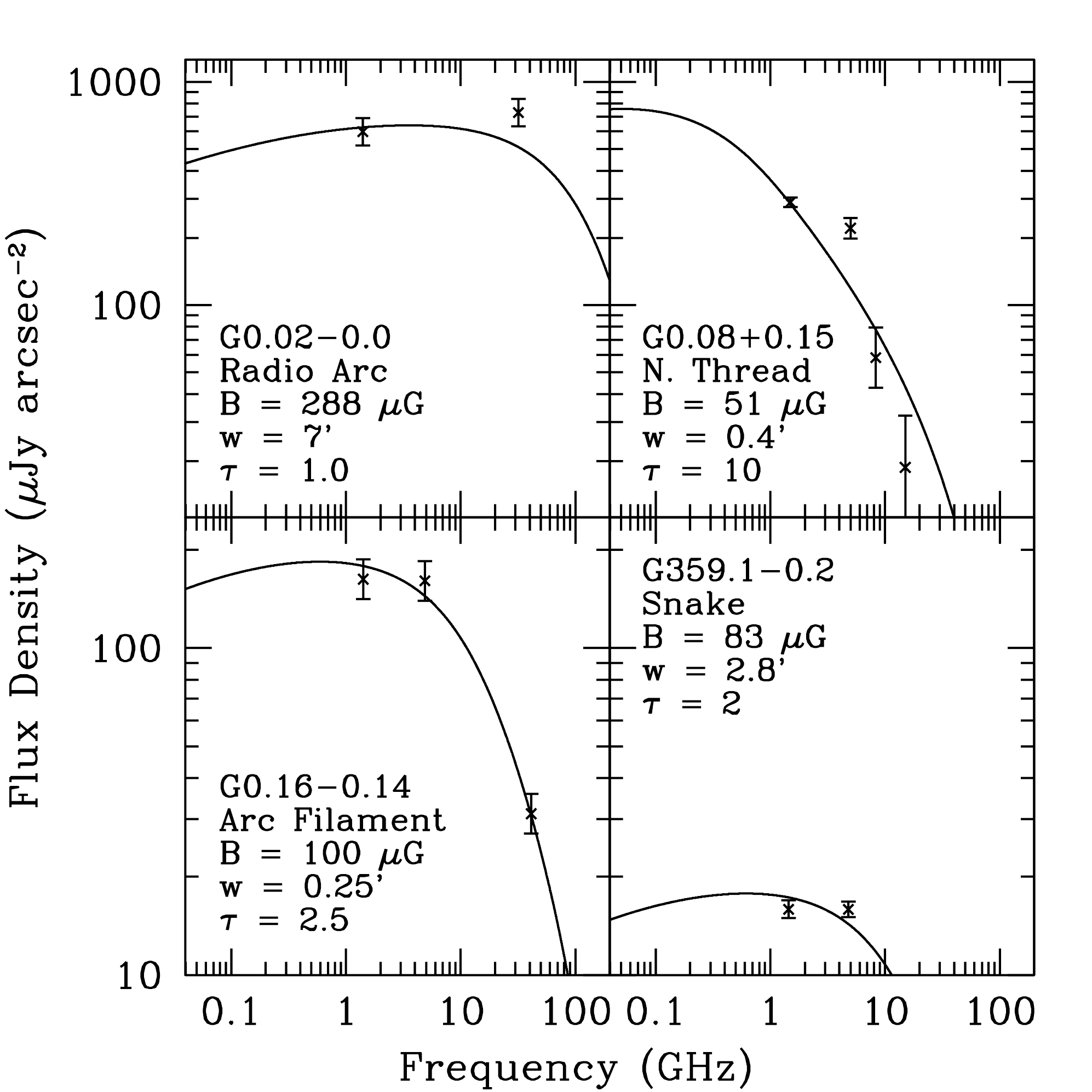}
		\caption{The synchrotron energy spectrum predicted from dark matter annihilations ($M_{\rm DM}=8$ GeV, annihilating to $e^+ e^-$, $\mu^+ \mu^-$ and $\tau^+ \tau^-$ with $\langle\sigma v\rangle=3\times 10^{-26}$ cm$^3$~s$^{-1}$) compared to the observed intensity and spectrum of G0.2-0.0 (the Radio Arc, top left), G0.08+0.15 (Northern Thread, top right), G0.16-0.14  (Arc Filament, bottom left) and G359.1-0.2 (the Snake, bottom right). The magnetic fields, filamentary width, and synchrotron energy loss times are shown for the synchrotron match to each filament.}
\label{fig:4}
\end{figure}

In Fig.~\ref{fig:4}, we provide dark matter fits to the intensity and spectrum of these four NRFs. For the Radio Arc (top left), we adopt a representative distance from the Galactic center of 20~pc, a diameter of 20~pc, a magnetic field of 290 $\mu$G and a diffusion timescale of 7.9~x~10$^{11}$~s ($\tau$~=~1.0). For the Arc Filament (bottom left), we adopt the same distance of 20~pc, using a filamentary diameter of 0.62~pc, a magnetic field of 100~$\mu$G and a diffusion timescale of 1.7~x~10$^{13}$~s ($\tau$~=~2.5). For the Northern Thread, we employ a Galactic center distance of 30~pc, a diameter of 1~pc, a magnetic field  of 50~$\mu$G, and a diffusion timescale of $2.5\times 10^{14}$~s. Finally, for the Snake we adopt a Galctic Center distance of 120~pc, a diameter of 7~pc, a magnetic field of 100~$\mu$G and a diffusion timescale of 1.9~x~10$^{13}$~s ($\tau$~=~2.0).

One potential mismatch in our best fit parameters concerns the larger widths necessary for the dark matter component to match the intensity of the observed radio filaments. While we have produced best fit widths of 7' for the Radio Arc, 0.4' for the Northern Thread, 0.25' for the Arc Filament, and 2.8' for the Snake, observational data supports smaller widths of 4' for the Radio Arc~\citep{reich2003}, 0.07'-0.2' for the Northern Thread~\citep{1999ApJ...526..727L}, 0.3' for the Arc Filament~\citep{1992PASJ...44..367S} and 0.2'-1.0' for the Snake~\citep{1995ApJ...448..164G,1991MNRAS.249..262A}. However this mismatch may be expected for three reasons. First, we have in general attempted to match the peak luminosities observed within the NRFs, while our simulations calculate the average luminosity expected throughout the enhanced magnetic field region of the NRF. Changes in the magnetic field intensity and structure may confine electrons more effectively in specific regions of a NRF, which would then show luminosities significantly brighter than those predicted by dark matter. Secondly, observations calculating the width of the NRFs would be expected to miss the outer regions of enhanced magnetic field, where the decreased width of the filament would produce only negligible emission. Third, the distance to the Galactic center may change considerably over the length of the NRF, producing enhanced annihilation rates at regions near the Galactic center. For instance, if the Northern Thread is more correctly modeled to extend linearly from 20~pc to 45~pc from the Galactic center, the total annihilation rate is enhanced by over 50\%. However, even these features may be insufficient to explain the extremely large diameters necessary to explain synchrotron emission in G359.1-0.2 (the Snake). We note that our analysis does not preclude astrophysical mechanisms formed within the filament from also contributing to the synchrotron emission spectrum, which may be the case in this particularly bright filament. 



A second issue in this analysis concerns the implication that the diffusion constant is significantly lower in filaments with particularly soft spectra. These disparities are not unique to a dark matter scenario, and have traditionally existed in astrophysical interpretations of these NRFs~\citep[cf.][]{1995PASJ...47..829T, 1999ApJ...526..727L}. In both dark matter and astrophysical scenarios, this may be understood in a model where the filaments exist as an entirely ordered magnetic enhancement superimposed on a random diffuse magnetic field of approximately 10~$\mu$G which permeates the Galactic center region. The differing ratios of the ordered to random magnetic fields (e.g 80\% in the Northern Thread vs. nearly 100\% in the Radio Arc) would then drive significantly enhanced parallel diffusion in the Radio Arc. Alternatively, assumptions that the Alfv\`en velocity places an upper limit on the speed of electron diffusion implies a diffusion timescale which scales as $B^{-1}$ and would approximately match the ratio of diffusion timescales observed in these two systems~\citep{1942Natur.150..405A}. We note, however, that this effect is not well understood and remains a significant assumption in our model. Lastly, it is possible that the magnetic field structures at the edges of the NRFs are configured to allow significant reflection of trapped electrons~\citep{1988ApJ...330..718H}.

Another necessary feature in any dark matter model of NRFs concerns the radial dependence of the electron injection spectrum. As shown in Eqn.~\ref{eq:annihilation_rate}, the dark matter annihilation rate within a given filament falls off as $\sim r^{-2.5}$, where $r$ is the distance of the filament to the dynamical center of the galaxy. A quantitative observation of the electron injection spectrum in individual filaments is difficult, due to the varying lengths, widths, magnetic fields, and diffusion constants in the observed filaments. However, the distance from the Galactic center to various NRFs is thought to span nearly an order of magnitude, which implies an injection spectrum that varies by more than a factor of 300 throughout the NRF population. This makes the statistical observation of such a feature possible, even with extremely crude estimations for the astrophysical parameters of individual NRFs. In order to examine this necessary trend, we have studied the observations of 7 NRFs with integrated fluxes and lengths observed at 330 MHz in the  \citet{2000AJ....119..207L} catalog, as well as the 13 NRFs observed at 1.4~GHz in the~\citet{2004ApJS..155..421Y} catalog. In both cases, integrated fluxes as well as lengths, are provided. We assume a constant radial width for all NRF, noting that quoted widths for most NRFs fall approximately within a factor of two. For this reason, we have removed the Radio Arc from our datasets as this assumption is particularly poor for that filament. 

The total luminosity of a NRF is expected to depend sensitively on its length. In addition to the linear dependence of the dark matter annihilation rate on the length of a filament, longer filaments are expected to retain electrons for longer periods of time and as a result will deposit a greater fraction of their initial energy into synchrotron radiation within the filament. In this work, we consider three scenarios to account for the influence of a NRF's length. First, we we consider the case in which electrons are effectively confined and lose their energy to synchrotron radiation on timescales much smaller than the diffusion timescale ($\tau\gg$~1). In this case the total flux in an NRF should depend only linearly on the length of the filament. Second, in the case that electrons free stream through the filaments on timescales much smaller than the synchrotron energy loss time ($\tau\ll$~1), the amount of energy deposited by a single electron into the filament is expected to scale with the length of the filament, providing a total flux which scales with the length of the filament squared. Finally, in the case that electrons diffusively propagate through the filament on a timescale smaller than the synchrotron energy loss time ($\tau \ll~$1 with D$_0$/c~$\ll$ filament length) the total energy deposited by an electron inside the filament will vary as the square of the filaments length, providing a total flux which varies as the cube of the length of the filament. The cases in which the total flux scales with $l$ and $l^{3}$ effectively bracket the possible degrees of correlation between the length of a NRF and it's total flux, while the $l^{2}$ case can be considered something of a median expectation.

\begin{figure*}
		\plotone{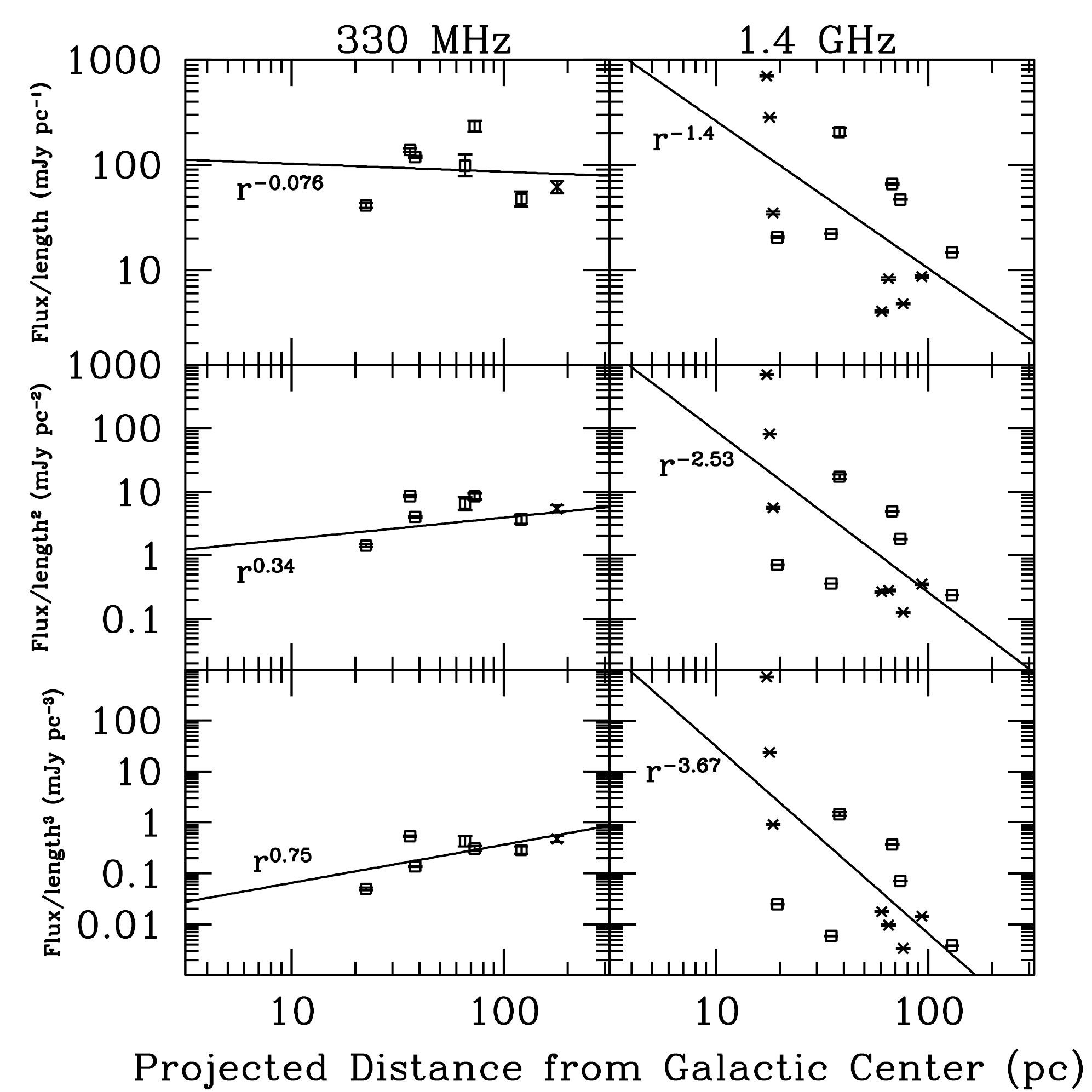}
		\caption{Flux per unit length (top), per length squared (center), and per length cubed (bottom) for NRFs at 330 MHz (left)\citep{2000AJ....119..207L} and 1.4 GHz (right) \citep{2004ApJS..155..421Y}, as a function of the projected distance of each filament from the Galactic center. Error bars based on the integrated flux, as well as a best-fit linear regression are shown for each frame. Boxed points indicate NRFs which are listed in both the 1.4 GHz and 330 MHz datasets. At 330 MHz, there is no clear correlation between the flux and the distance of a filament from the Galactic center.  At 1.4 GHz, however, those filaments closer to the Galactic center are clearly brighter than those farther away.}
\label{fig:5}
\end{figure*}

 We first examine the observed dataset at 330~MHz. In the left frames of Fig.~\ref{fig:5}, we plot the flux per unit length (top), per unit length squared (middle) and per unit length cubed (bottom) as a function of the projected distance of each NRF to the Galactic center. In each case, we note no significant trend between the distance of a given filament from the Galactic center. In other words, the distance of a given filament from Galactic center does not appear to have significant bearing on its emission at 330~MHz, suggesting that astrophysical mechanisms ({\it i.e.} not dark matter annihilations) are responsible for the emission at this frequency. 
 
The same conclusion is not found at 1.4~GHz, however. At this frequency (right), we see a very significant correlation between the projected distance of a filament to the Galactic center and its observed intensity. In particular, filaments closer to the Galactic center tend to be considerably brighter at 1.4~GHz than those farther away. We note that for the dark matter halo profile used in this paper, we predict a flux which scales with $r^{-2.5}$, while a more generic range of profiles predicts behavior between roughly $r^{-2}$ and $r^{-3}$.

There are several interesting features of the results shown in Fig.~\ref{fig:5}. First, although the correlation observed among the filaments in the 1.4 GHz dataset could have plausibly been the result of a selection effect resulting from the presence of greater astrophysical backgrounds closer to the Galactic center, the 330 Mhz observations dispute this reasoning, as does the lack of any bright NRFs at 1.4 GHz far from the Galactic center. The different trends observed in these datasets instead support the surprising conclusion that the electron injection spectra which produces the emission at 1.4 GHz is physically distinct from the emission mechanism which dominates at 330 MHz. Both trends are naturally explained in the case of a light dark matter particle in a magnetic field of $\sim$ 100~$\mu$G, as the synchrotron spectrum will peak at approximately 1~GHz, with astrophysical processes dominating the emission at lower frequencies.

We caution that this relationship is still somewhat tenuous, due to the varying NRF widths, magnetic fields, three dimensional distances, and diffusion constants which are not modeled in this analysis, and additional studies will be necessary to better identify and refine this apparent trend. If dark matter annihilations are in fact responsible for a sizable fraction of the emission observed at 1.4 GHz, improved measurements of these parameters should enhance the correlation shown in Fig.~\ref{fig:5}. Furthermore,  if astrophysical mechanisms dominate the low energy synchrotron emission, then we would expect the spectral slope in this region to be relatively soft. We note that the observed trend is steepest among the 1.4 GHz NRFs which were not included in the 330 MHz analysis, making additional measurements in this region necessary to better understand the apparent mismatch between low and high frequency observations. Finally, we note that no correlation such as that described here would be expected in reconnection or shock acceleration models of NRF. However, models based on a monoenergetic electron flux from Sgr A* could plausibly lead to a similar relation, although as argued in \S\ \ref{sec:filamentaryarcs}, such electrons are not expected to penetrate into the NRFs.

\section{Discussion and Conclusions}
\label{sec:conclusion}

The observed synchrotron emission from non-thermal radio filaments (NRFs) in the Inner Milky Way have long been difficult to explain with known astrophysical mechanisms. In this article, we have proposed that dark matter annihilations taking place within these filaments could produce the nearly monoenergetic electron spectrum necessary to generate the hard synchrotron emission that is observed. In this regard, dark matter annihilations have several advantages over proposed astrophysical mechanisms. First, electrons produced through dark matter annihilations yield a synchrotron spectrum in good agreement with the hard spectral index and turnover observed from NRFs, without mandating finely tuned magneto-hydrodynamic interactions to move electrons independently into equivalent monoenergetic electron distributions. Second, WIMPs annihilating into leptonic final states, such as those employed at low masses in order to explain the Fermi-Galactic center excess~\citep{2011PhLB..697..412H}, are predicted to inject electrons into the Galactic center region with an energy density very similar to the synchrotron signal observed from NRFs. Lastly, as shown in Fig.~\ref{fig:5}, dark matter annihilations naturally explain the observed correlation between the radial distance of NRFs to the Galactic center and the inferred electron injection spectrum in the filaments.

A dark matter origin of the observed radio emission from NRFs yields several concrete and testable predictions. In particular, the dark matter annihilation rate and corresponding flux of injected electrons must show approximate spherical symmetry with respect to the Galactic center~\citep{2008Natur.454..735D}, and the injected spectrum of electrons must be identical for all filaments throughout the Galactic center. This is not the case for other proposed astrophysical mechanisms, for which the electron injection spectrum can vary from filament to filament. Furthermore, we have demonstrated that the intensity of 1.4 GHz emission is significantly enhanced for NRFs near the Galactic center compared to those farther away, as would be expected from dark matter distributed in a cusped halo profile. Observations which are able to independently determine or constrain the magnetic field strengths and other characteristics among a population of NRFs could be used to further examine any scenario requiring a single electron injection spectrum. If dark matter annihilations are found not to power NRFs, these objects may be used to place stringent constraints on the dark matter annihilation rate in the region surrounding the Galactic center. New observations of NRFs will thus be integral to resolving both the astrophysics and particle physics of these unusual regions.

\acknowledgements
We would like to thank Pasquale Blasi, Greg Dobler, Olindo Dumbser, Doug Finkbeiner, Iris Gebauer, Stefano Profumo, and Andy Strong for helpful comments and discussions. TL is supported by a Fermilab Fellowship in Theoretical Physics. DH is supported by the US Department of Energy and by NASA grant NAG5-10842. 

\bibliography{filaments}

\end{document}